\documentstyle[psfig,conf_iap]{article}
\begin{document}
\heading{%
%Begin Heading
%
Submillimetre-wave surveys: first results and prospects
%
%End Heading
} 
\par\medskip\noindent
\author{%
%Begin Author names
A.\,W. Blain$^1$
%End Author names
}
\address{%
%First address
Cavendish Laboratory, Madingley Road, Cambridge, CB3 0HE, UK. 
}

\begin{abstract}
The population of distant dusty submm-luminous galaxies was first 
detected last year \cite{SIB}. Forms of evolution required to account for both
this population and the intensity of background radiation have now been 
determined \cite{BSIK}, and are used to investigate the most efficient 
observing strategies for future surveys. 
\end{abstract}

\medskip

Submm-wave galaxy surveys detect the redshifted thermal far-infrared 
radiation emitted by dust grains that reprocess optical/ultraviolet light from 
young stars and active galactic nuclei (AGN). The selection function in redshift in
such surveys is very broad \cite{BL}, and extends out to redshifts of order 10. 
Hence they provide an efficient and direct technique for selecting 
distant dust obscured galaxies and AGN \cite{I+7}. High-redshift biased 
selection makes submm-wave surveys an ideal way to search for gravitational 
lenses \cite{B}.

Consistent models of the evolution of dusty galaxies can be constructed to 
account for submm counts \cite{Barger, Eales, Holland, Hughes, Lilly, SIB}, 
mid-infrared counts \cite{Kawara, Lagache} and the 
submm/far-infrared background radiation intensity \cite{Fixsen, Hauser, Puget, 
SFD}. While not necessarily correct, these models are well constrained and can 
be used to predict source counts on a range of different angular scales and at a
range of different wavelengths \cite{BSIK}. Here the resulting counts are used 
to investigate the optimal strategies for future mm/submm-wave surveys. The 
source confusion noise expected in these surveys is discussed 
elsewhere \cite{BIS, BISK}. 

Observed submm-wave counts are presented in Fig.\,1, alongside the counts 
predicted by taking an ensemble average of seven different models of galaxy
evolution that are consistent with both the observed counts and the intensity of 
background radiation \cite{BSIK}. The associated redshift distributions are 
shown in Fig.\,2. Excluding \cite{I+7}, no redshifts have yet been determined for 
submm-selected sources. A spectroscopic redshift distribution of 
submm-selected sources will provide a powerful test of plausible galaxy 
evolution models \cite{BSIK}. The 
broad-band colours of the optical identifications in general suggest agreement 
with the predictions \cite{SIBK}.

\begin{figure}[t]
\psfig{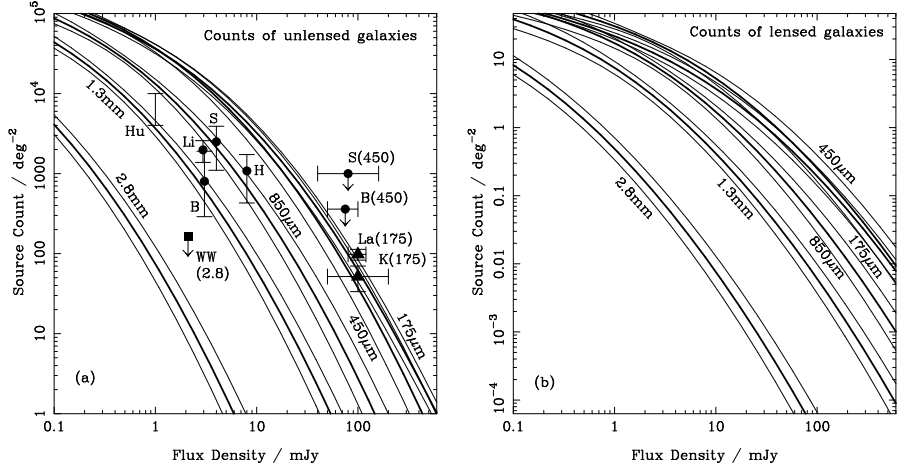}
\vskip -0.35cm
\caption[]{Observed source counts and the ensemble mean (thick lines) and 
1$\sigma$ uncertainty (thin lines) derived from well-fitting count and
background models \cite {BSIK}. References to the data are B\cite{Barger}, 
H\cite{Holland}, Hu\cite{Hughes}, K\cite{Kawara}, La\cite{Lagache},
Li\cite{Eales,Lilly}, 
S\cite{SIB} and WW\cite{WW}. Points without a number correspond to 
850-$\mu$m data. 
}
\vskip -0.35cm
\end{figure}

\begin{figure}[t]
\psfig{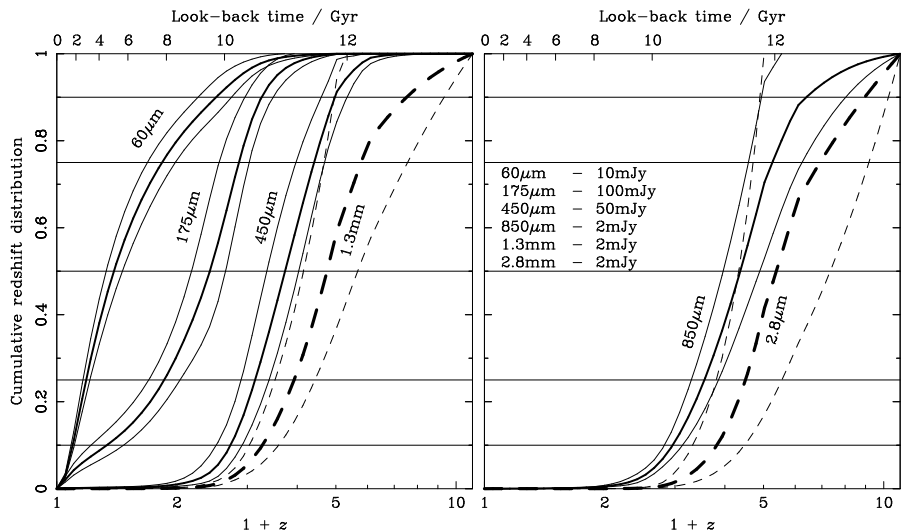}
\vskip -0.35cm
\caption[]{The ensemble mean (thick lines) and 1$\sigma$ uncertainty (thin 
flanking lines) of predicted submm-selected redshift distributions \cite{BSIK}. 
}
\vskip -0.35cm
\end{figure}

The predicted counts of mm/submm-wave sources can be used to determine the 
most efficient strategy for detecting distant dusty galaxies, in both general galaxy 
surveys (Fig.\,3a) and in surveys for lensed objects (Fig.\,3b). The detection rate 
achieved in the crucial first detections using SCUBA will be greatly exceeded 
in future surveys; for references to suitable instruments see \cite{BIS}. 
The fraction of lensed sources expected in a survey is shown in 
Fig.\,3(c). The rate of {\it confirmation} of lenses, including the time required to 
image lensed structures at sub-arcsec resolution using the MMA, is shown in 
Fig.\,3(d). The detection and confirmation rates differ most at faint flux 
densities, at which the follow-up MMA observations require more time.

\begin{figure}[t]
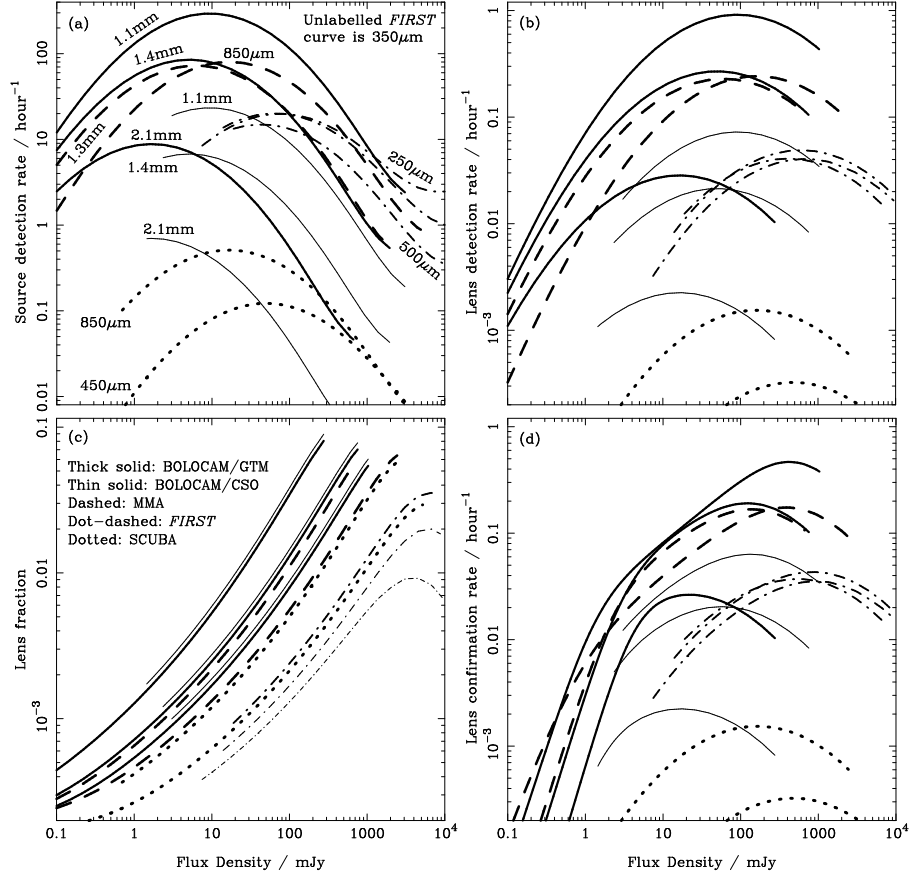

\psfig{figure=blaincontribF3a.ps,width=5.8cm,angle=-90}
\vskip 0.05cm \hskip 0.05cm
\psfig{figure=blaincontribF3b.ps,width=5.9cm,angle=-90}
\vskip -3.5mm 
\caption[]{Predicted detection rates of unlensed (a) and lensed (b) 
mm/submm-wave sources at 5$\sigma$ significance. The rates are uncertain to 
within a factor of about 3. The curves end on the left at a flux density 5 times the 
confusion noise \cite{BISK} and on the right at a count of 1/4$\pi$\,sr$^{-1}$. The 
fraction of lensed sources (c), and the {\it confirmation} rate of lenses (d), after 
MMA follow-up observations, are also shown. The fraction of lenses is expected 
to increase at bright flux densities, and to be systematically larger at longer 
wavelengths. Although the surface density of sources is expected to be small 
when the lens abundance is greatest, a detection efficiency greater than one per 
cent should still be achieved in a deeper survey. Lens confirmation rates in 
(d) were calculated assuming that a 850-$\mu$m MMA observation 10 times 
deeper than the detection threshold is required in order to search for lensed 
structures in each source. At faint flux densities the time required for these 
follow-up observations greatly exceeds the time required to carry out the initial 
survey. Lens surveys should hence be conducted at brighter flux density 
limits than galaxy surveys. 
}
\vskip -0.35cm
\end{figure}

The detection rates predicted for future instruments \cite{Glenn} and telescopes 
-- an upgraded `SCUBA+' \cite{HollandPC}, large ground-based interferometers 
like the MMA, the 50-m LMT, a 10-m South Pole telescope and the 3.5-m 
space-borne {\it FIRST} -- exceed those of SCUBA by up to two orders of 
magnitude. In concentrated efforts over a number of years, catalogues of order 
10$^6$ distant galaxies and AGN will be compiled. The fine angular resolution of 
large interferometers will allow the detected sources to be resolved directly 
in the mm/submm waveband. In some cases, their redshifted mid-infrared 
fine-structure line emission could be used to determine redshifts directly in the 
submm waveband.

In addition, the cosmic microwave background imaging space mission {\it Planck 
Surveyor} will provide an all-sky arcmin-resolution 100-mJy survey, which will 
be very useful for selecting both extremely luminous submm-wave sources 
and gravitational lenses \cite{B}. 

\begin{itemize}
\item The first detections of distant dusty galaxies indicate that there is an 
abundant population of such sources. These can be exploited to investigate 
galaxy formation and evolution, large-scale structure at high redshifts and 
the values of cosmological parameters.
\item It is most important to determine the redshift distribution of submm-wave 
sources selected in blank-field surveys, and thus test the first observationally 
constrained models of the evolution of dust obscured galaxies at high
redshifts \cite{BSIK}.
\item Huge samples of distant dusty galaxies and gravitational lenses can be 
compiled using future instruments -- especially the LMT and MMA.
\end{itemize}

\acknowledgements{This work has benefited greatly from SCUBA/JCMT 
observations in collaboration with Ian Smail, Rob Ivison and 
Jean-Paul Kneib. I thank Malcolm Longair for helpful comments on the 
manuscript, and Gislaine Lagache and Dave Clements for discussing the 
results of the {\it ISO} FIRBACK programme.}

\vfill
\end{document}